\documentclass[runningheads]{llncs}
\usepackage[T1]{fontenc}
\usepackage{graphicx}
\usepackage{booktabs} 
\usepackage{array}    
\usepackage{caption}  

\usepackage{graphicx} 
\usepackage{subcaption} 
\usepackage{amsmath} 

\begin{document}
\title{Gender Attribution in Causal Beliefs}
%
\author{{Anonymous Submission}}
\author{Zhuoyu Shi\inst{1, 2}\orcidID{0009-0009-7936-3931} \and
Fred Morstatter\inst{1,2}\orcidID{0000-0002-0247-4328} }
\authorrunning{Shi and Morstatter}
%
\institute{Thomas Lord Department of Computer Science, University of Southern California, Los Angeles, CA 90089, USA \and
Information Sciences Institute, University of Southern California, Marina del Rey, CA 90292, USA
\email{\{zhuoyush,morstatt\}@usc.edu} }
%
\maketitle              
\begin{abstract}
For centuries, women have been cast as the source of harm in public narratives, from witch hunts in early modern Europe to contemporary stereotypes about emotional instability. These cultural patterns reflect enduring biases in how people attribute causality and assign blame, often portraying women as agents of disruption and men as figures of rational authority. In this study, we examine how such gendered causal attributions appear in everyday language. Leveraging three complete 24-hour datasets of all English-language posts on Twitter, and using language models, we extract cause-and-effect relationship pairs and identify gendered attribution of causal agents. We then analyze how gender attribution relates to sentiment, the kinds of effects invoked, and the diffusion of posts through the social networks. Our findings reveal that female-attributed causes are more often associated with negative sentiment and emotional or relational outcomes, whereas male-attributed causes are more frequently linked to positive sentiment and abstract, structural effects. Moreover, male-attributed narratives spread more widely across communities. These results suggest that longstanding gender stereotypes continue to appear in how people express and amplify causal narratives in public discourse, in decentralized, high-velocity environments like social media.

\end{abstract}

\section{Introduction}

Across centuries of recorded history, gender has been a powerful axis of social attribution, shaping how individuals are blamed, credited, and understood in the narratives that define public life. Women, in particular, have long been cast as emotional, unstable, or morally suspect actors in dominant cultural frames. From the witch hunts of early modern Europe, where women were disproportionately accused of supernatural harm~\cite{levack2013witch}, to the persistent stereotype of the ``irrational'' woman in political rhetoric~\cite{faludi2009backlash}, gendered causal attribution has been a consistent mechanism of control and marginalization. These patterns are not just vestiges of history; they continue to operate in modern discourse, subtly influencing how actions and consequences are narrated, interpreted, and shared.

Social science research has shown that gender stereotypes organize how we assign agency, intent, and consequence~\cite{eagly1984gender,rudman2001implicit,fiske2018model}. Women are more likely to be described in terms of emotions, relationships, or interpersonal disruption, while men are often portrayed as rational actors, leaders, or structural influencers. Language plays a central role in this process. Yet, the role of gender in causal framing remains underexplored. The way people use causal language, how they construct ``who did what'' and ``what happened as a result'' offers a valuable window into the collective beliefs and biases that structure public narratives.

In this study, we investigate how gender is attributed within human causal beliefs, using the linguistic structure of causal language as a lens. Social media offers a unique environment for this inquiry: it contains vast volumes of spontaneous, high-frequency discourse, in which people narrate events, make attributions, and express judgments in real time. To this end, we analyze a complete set of messages from a 24-hour period on Twitter~\cite{pfeffer2023just}, and validate our findings with two additional days of full-platform data. This large-scale analysis allows us to examine how gender attribution manifests in everyday causal language, and how it interacts with sentiment, narrative content, and information diffusion.

Whereas prior studies have focused on explicit gender bias, such as visibility gaps or abusive language, our research turns attention to a subtler but equally consequential dimension: the use of causal assertions~\cite{eagly1984gender,rudman2001implicit,fiske2018model,shi2024diffusion}. Specifically, we aim to understand how gender influences the construction and spread of causal narratives on social platforms. We address the following research questions:

\begin{itemize}
    \item \textbf{RQ1} – How do sentiments differ across different gender attributions in tweets?
    \item \textbf{RQ2} – How do the types of effects attributed to females differ from those attributed to males?
    \item \textbf{RQ3} – How does engagement vary based on different gender attributions within tweets?
\end{itemize}

To answer these questions, we first extract causal language by identifying cause-and-effect span pairs from tweets. We then identify whether the cause expression contains an explicit gender reference, validating our approach using human annotations from a controlled labeling task. With this annotated dataset, we examine how sentiment varies across gendered causes, analyze the types of effects associated with each gender, and assess differences in engagement and propagation across communities.

Our findings reveal consistent asymmetries in how gender is expressed, framed, and diffused through causal language. Female-attributed causes are more likely to be expressed with negative sentiment and linked to emotional or relational consequences, while male-attributed causes tend to be framed more positively and are associated with structural or abstract effects. Moreover, male-attributed narratives propagate more widely within and across communities. These findings shed light on how deep-seated gender norms continue to shape public discourse, even in informal, decentralized communication environments. Understanding these dynamics is critical for developing more equitable approaches to information analysis, media literacy, and platform governance.

\section{Related Work}

Gender bias has long been a central focus of social science research, particularly in how individuals perceive and evaluate agency, emotion, and responsibility. A robust body of work in psychology and sociology has shown that gender stereotypes shape expectations about who causes what, how their actions are interpreted, and what consequences are imagined. Women are often associated with communal traits such as warmth, empathy, and emotionality, while men are more closely linked to agentic qualities such as competence, control, and assertiveness~\cite{eagly1984gender,fiske2018model,ridgeway2001gender,rudman2001implicit}. These stereotypes influence not only interpersonal judgment but also broader patterns of discourse, including how people attribute blame, assign credit, and frame social events~\cite{heilman2001description,hastorf1954they,meehan2012gendering}.

Narrative and linguistic framing are particularly important mechanisms through which gender bias operates. Decades of research in sociolinguistics and discourse analysis have shown that women are more likely to be described in affective, relational, or interpersonal terms, while men are more often framed in relation to structure, action, and abstraction~\cite{lakoff1973language,tannen1990you,coates2015women,kitzinger1999just}. These discursive patterns shape not only how stories are told but also whose voices and perspectives are perceived as credible or authoritative. Prior work shows that even subtle linguistic choices can reinforce unequal power dynamics, for instance, by associating women with emotional labor or vulnerability and men with leadership or rationality~\cite{brescoll2008can,glick2018ambivalent}. When causal language reflects these biases, it contributes to the reproduction of social hierarchies by naturalizing different expectations for men and women in public narratives.

Recent work in computational social science has extended these insights by demonstrating that such gendered patterns are measurable at scale. From sentiment distributions and topical framing to the frequency of gendered nouns and the structure of attribution, large-scale linguistic data consistently reveals asymmetric patterns that mirror long-standing social biases~\cite{bolukbasi2016man,mehrabi2021survey,stanczak2021survey}. However, while prior studies have examined representation, sentiment, or visibility in isolation, relatively few have explored how gender shapes causal explanations: who is framed as the source of an action or event, and how those frames differ in emotional tone and social consequence. This study builds on and extends these literatures by systematically analyzing gender attribution in cause-effect structures, revealing how inequality is embedded in the very architecture of explanation.

 \section{Data Description}

This study uses data collected in prior work that captured every public tweet posted during a 24-hour period on Twitter \cite{pfeffer2023just}. The primary dataset spans from September 20, 2022, 15:00:00 UTC to September 21, 2022, 14:59:59 UTC, comprising approximately 375 million tweets in total.

To focus on original user-generated content and minimize noise, we restrict our analysis to original (i.e., not a repost, reply, or quote post) English-language tweets that do not contain hashtags, URLs, or user mentions. This filtering ensures a cleaner textual context for linguistic analysis and reduces the influence of automated or promotional content, which often includes such features.

An initial propagation analysis revealed that the vast majority of retweets occur within 21.24 minutes of the original tweet’s posting time. To ensure that all posts in our sample have sufficient time to be reposted, we exclude tweets from the final hour of the 24-hour window. Our final dataset thus includes 6,158,756 original English tweets (excluding those with hashtags, URLs, or user mentions) from the first 23 hours of the day, of which 515,922 received at least one retweet within the full 24-hour period.

To assess the robustness of our findings, we repeat our analysis on two additional full-day Twitter datasets collected using the same methodology \cite{pfeffer2023just}. The first spans December 7, 2022, 11:00:00 UTC to December 8, 2022, 10:59:59 UTC, and the second spans January 25, 2023, 14:00:00 UTC to January 26, 2023, 13:59:59 UTC. From each, we again extract original English tweets (excluding those with hashtags, URLs, or user mentions) from the first 23 hours. The December 8 dataset includes 6,183,515 such tweets, of which 474,819 received at least one retweet. The January 26 dataset includes 5,459,854 such tweets, with 438,011 retweeted at least once within the full 24-hour window.

\begin{table*}[t]
\centering
\small
\begin{tabular}
{>{\centering\arraybackslash}p{1.60cm} >{}p{0.42\textwidth} >{}p{0.12\textwidth}  >{}p{0.12\textwidth} >{}p{0.06\textwidth}  }

\toprule

\multicolumn{1}{c}{\centering}{} & \multicolumn{1}{c}{\centering Tweet} & \multicolumn{1}{c}{\centering Cause}  & \multicolumn{1}{c}{\centering Effect} & \multicolumn{1}{c}{\centering Gender Attribution}\\

\midrule

Example 1 & It's my love from my grandmother make me gentle when I care for you & my grandmother & my love & Female \\
\addlinespace[2.5pt]

Example 2 & Blame my trust issues on my mom.  She used to threaten me and never follow through. & my mom & my trust issues & Female  \\
\addlinespace[2.5pt]

Example 3 & Still getting surprise snacks from my dad when he makes random trips to the grocery & my dad & surprise snacks & Male  \\
\addlinespace[2.5pt]



\bottomrule
\end{tabular}
\caption{Examples of tweets, cause-and-effect pairs, and gender mentioned in causes. }
\label{tab:CF_pairs}

\end{table*}

\section{Computational Methods}

\subsection{Detecting Cause-and-Effect Span Pairs from Text}

To identify causal claims made in user-generated text, we employ a state-of-the-art deep learning model~\cite{priniski2023pipeline}, which has been fine-tuned based on RoBERTa~\cite{liu2019roberta}, a transformer-based language model. it extracts the specific spans corresponding to the ``cause" and the ``effect." This span-level granularity is essential for our research goals, as it allows us to isolate the precise linguistic framing users employ when attributing causality, particularly useful when analyzing sentiment or gender reference in causal expressions. This model reaches an F1 score of 0.874, where the precision rate is 0.883, and the recall rate is 0.865~\cite{priniski2023pipeline}.

\subsection{Gender Attribution in Causes}

A central objective of this study is to examine how gender is implicated in the framing of causal statements on social media. Specifically, we are interested in understanding whether users attribute different types of outcomes, sentiments, or social consequences depending on the gender of the individual or group implicated in a causal role. Focusing on gender mentions in the cause rather than in the effect allows us to anchor our analysis on the agents or sources of actions and outcomes. To systematically investigate how gender attribution shapes causal framing and sentiment on social media, it is essential to first identify whether a given cause expression explicitly references a gendered subject. In the context of our study, this enables us to isolate and compare how actions, sentiments, and effects differ when users attribute causality to individuals perceived as male or female. Without this disambiguation, it would be difficult to draw meaningful conclusions about gendered patterns in causal language, engagement, or affective framing.

To this end, we employ the GPT-4o language model to automatically annotate the gender mentioned in the cause portion of each tweet with `Male', `Female', or `None'.

\subsection{Validation of Gender Annotation}

We conduct a human evaluation study to validate the reliability of our GPT-4o-based gender annotation pipeline. We randomly sampled a total of 200 messages from our dataset: 100 from the subset annotated by the model as \textit{Male} and 100 from the subset annotated as \textit{Female}.

Human annotations were collected via the Prolific platform. Annotators were first trained with five labeled examples. They were then evaluated in four test cases. We only included annotators who correctly annotated at least three out of four of these examples. Each qualifying annotator annotated 25 messages and received a compensation of \$3.

Each tweet was independently labeled by three distinct annotators with one of the options: \textit{Male}, \textit{Female}, or \textit{Other}. Final annotations were determined via majority vote to ensure consistency and mitigate individual annotator bias.

The GPT-4o model achieved a precision of 91\% for male attributions and 93\% for female attributions. To evaluate the consistency among annotators, we then calculate the inter-annotator agreement with Fleiss' Kappa. The overall agreement is 85\%. These results indicate that the model performs reliably in identifying explicit references to gender within the cause portion of the tweets. The high precision across both categories supports the use of GPT-4o as an effective tool for gender annotation in causal language contexts.

\section{RQ1 - Sentiment Differences Across Gender Attributions}

To examine whether sentiment expression on Twitter varies by gender attribution in causal narratives, we first classified the sentiment of each tweet using the VADER sentiment analyzer \cite{hutto2014vader}. VADER is a lexicon-based tool specifically designed to capture sentiment in informal, short-form text, making it well suited for social media analysis. Each tweet was automatically labeled as \textit{positive}, \textit{neutral}, or \textit{negative} based on VADER's compound score thresholds.

We then grouped tweets by the gender attributed in the causal clause: specifically, whether the subject of the cause was identified as male or female. Only tweets where gender was explicitly labeled as Male or Female were included in this analysis. For each day in our three-day sample window, we computed the proportional distribution of sentiment categories within the male- and female-attributed tweet groups.

\begin{figure}[tbhp]
\centering
\includegraphics[width=.80\linewidth]{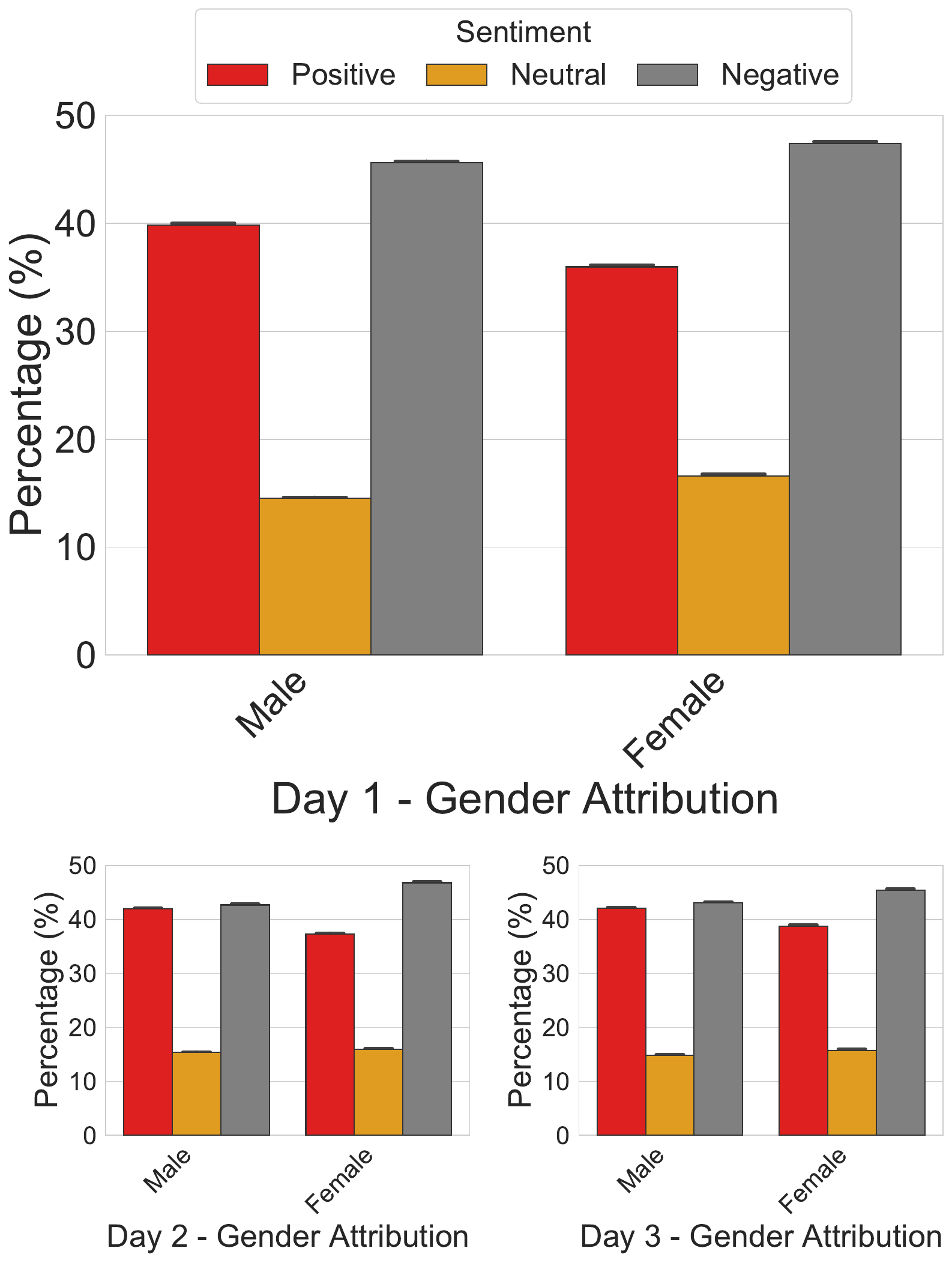}
\caption{Distribution of sentiment across male- and female-attributed tweets over three sampled days (100 bootstrapped runs). Female-attributed causes are consistently associated with higher proportions of negative sentiment, while male-attributed causes show higher positive sentiment.}
\label{fig:RQ1}
\end{figure}

Our findings, shown in Figure~\ref{fig:RQ1}, reveal consistent sentiment asymmetries across all three days. Tweets in which causality is attributed to female subjects show a significantly higher proportion of negative sentiment compared to those attributed to male subjects. In contrast, male-attributed tweets show slightly higher rates of positive sentiment. Neutral sentiment distributions remain relatively stable.

These patterns suggest that public discourse on Twitter does not express gendered causal narratives equally in affective tone. Instead, gender attribution appears to differ by the emotional framing of events: female subjects are more frequently linked with negative emotional language. It reflects known biases in human perception and attribution, where women are more likely to be viewed as responsible for interpersonal or affective disruption, while men are framed as agents of competence or authority \cite{eagly1984gender,fiske2018model}.

Importantly, these sentiment asymmetries emerge in a context where language is already constrained to brief, high-signal formats. That they persist even in short, causal tweets suggests that gender bias is  embedded in the everyday, fast-paced narrative structures that dominate digital communication. The observed differences in sentiment are subtle at the aggregate level but statistically consistent, raising concerns about the effects of implicit bias on Twitter.

\section{RQ2 - Types of Effects Attributed to Different Gender}

To explore how the \textit{types of effects} attributed to female and male subjects differ, we analyze the nouns that appear as effects in extracted cause-effect pairs, conditioned on the presence of a gender attribution in the cause. We further stratify this analysis by sentiment polarity (positive vs. negative) to understand whether affective framing modulates these differences.

For each noun identified as an effect, we compute an \textit{adjusted frequency} to normalize across the unequal number of gender-attributed tweets. Specifically, for a given noun, its frequency in female-attributed effects is scaled by the relative proportion of female-attributed tweets:

\[
\text{AdjustedFreq}_{\text{female}} = \text{RawFreq}_{\text{female}} \times \frac{N_{\text{female}} + N_{\text{male}}}{N_{\text{female}}}
\]

where $\text{RawFreq}_{\text{female}}$ denotes the word frequency within effects that causes contains a reference to female, and  \( N_{\text{female}} \) and \( N_{\text{male}} \) denotes the total number of tweets containing gender attribution, respectively. The same adjustment is applied to male noun frequencies. This ensures that comparisons reflect relative association strength rather than raw counts.

\begin{figure*}
    \centering
    \begin{subfigure}{0.9\textwidth}
        \centering
        \includegraphics[width=0.93\linewidth]{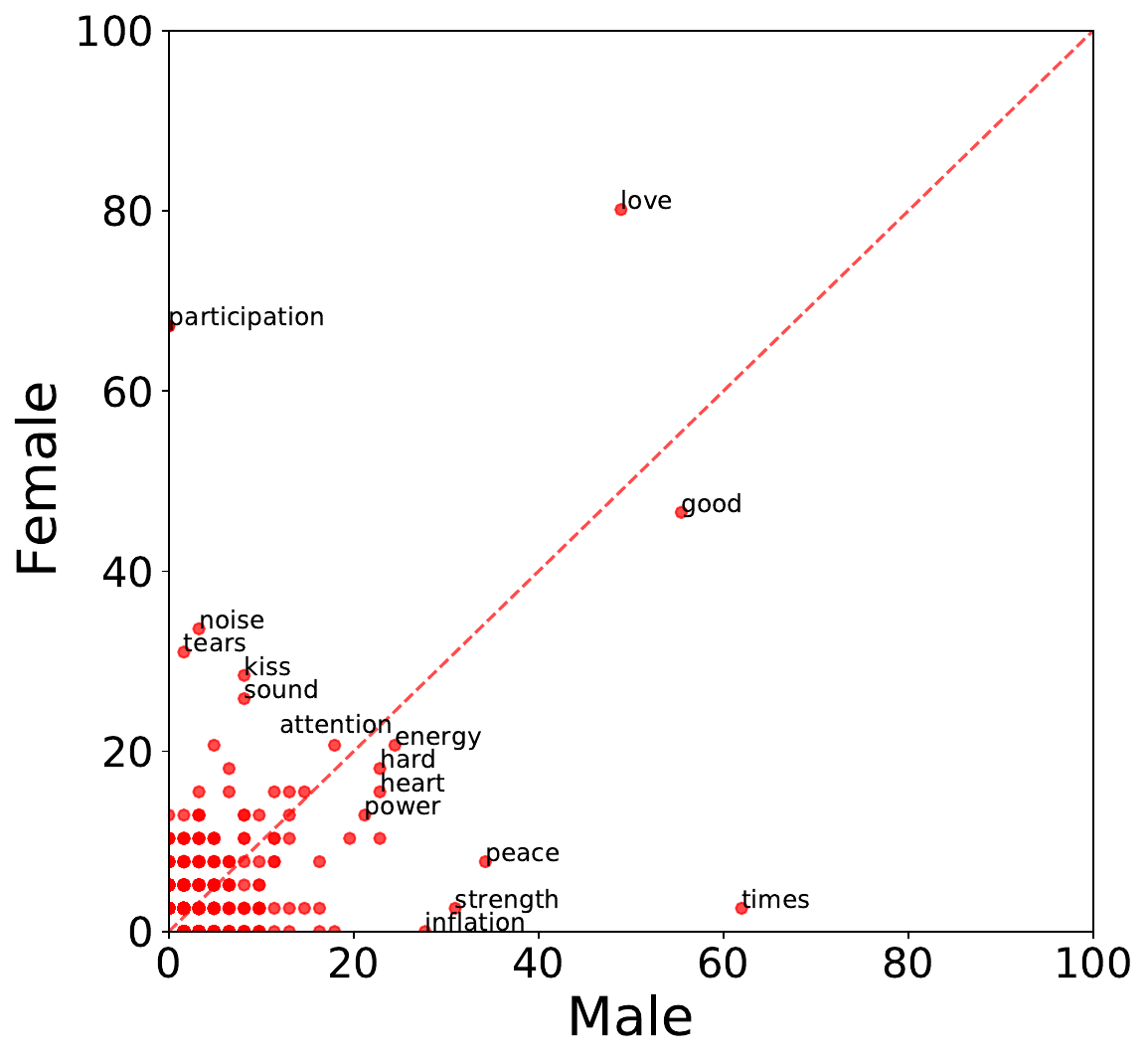}
        \caption{Positive}
        \label{fig:RQ2_pos}
    \end{subfigure}
    \begin{subfigure}{0.9\textwidth}
        \centering
        \includegraphics[width=0.93\linewidth]{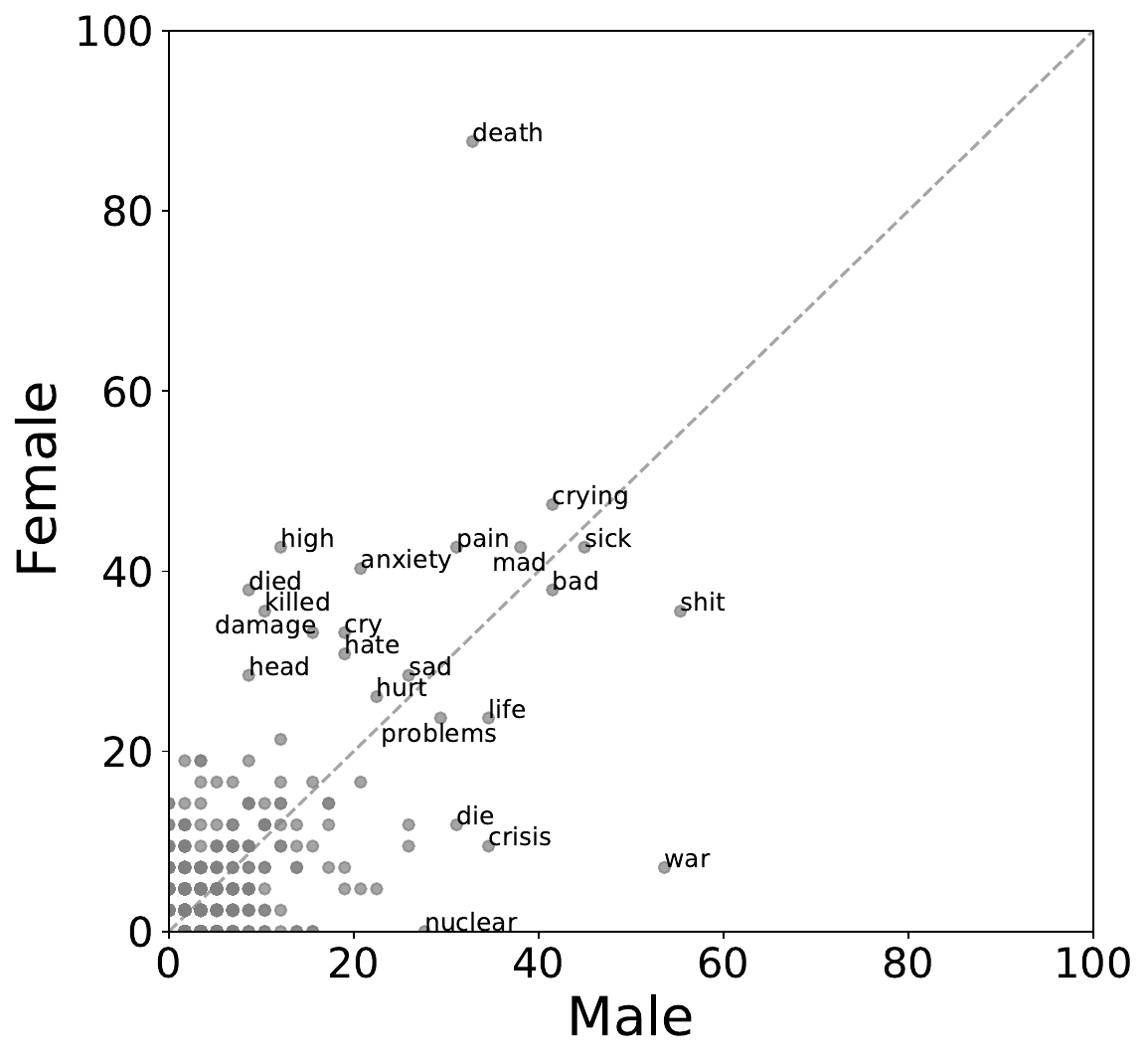}
        \caption{Negative}
        \label{fig:RQ2_neg}
    \end{subfigure}

     \caption{Gendered distribution of effect nouns in cause-effect pairs under (a) positive and (b) negative sentiment. Axes indicate adjusted frequencies.} 
    \label{fig:RQ2}   
\end{figure*}

Figure~\ref{fig:RQ2} visualizes the resulting distribution of effect nouns for positive sentiment (left panel) and negative sentiment (right panel). Each point corresponds to a noun, with its $x$-axis and $y$-axis values representing the adjusted frequencies in male- and female-attributed tweets, respectively. The diagonal line represents parity: nouns above the line are more frequently associated with female attributions, and those below the line are more male-associated.

The effects associated with female attributions tend to be more relational and emotive, including nouns such as \textit{love}, \textit{participation}, \textit{attention}, and \textit{tears}. In contrast, male-attributed effects skew toward abstract or structural nouns like \textit{strength}, \textit{power}, \textit{inflation}, and \textit{peace}. This suggests that in positively framed discourse, female subjects are more often credited with or linked to interpersonal or emotional outcomes, whereas male subjects are associated with systemic or impersonal outcomes.

Under negative sentiment, we observe a similarly gendered pattern but in reverse valence. Female attributions are more frequently connected to internalized or affective states such as \textit{crying}, \textit{pain}, \textit{anxiety}, and \textit{death}. Male attributions, on the other hand, are disproportionately associated with broader-scale or external consequences, such as \textit{war}, \textit{nuclear}, and \textit{shit}. These divergences suggest that causal language on Twitter reflects stereotyped associations between women and emotional distress, and between men and societal-scale disruption or aggression.

The observed asymmetries in gendered attribution of effects reveal consistent patterns that align with prior work in sociolinguistics and computational social science. Specifically, female-attributed effects are more likely to be emotionally or relationally focused, while male-attributed effects emphasize structural, external, or forceful consequences. This resonates with long-established findings that language about women tends to foreground affect, vulnerability, and interpersonal dynamics, whereas language about men more often emphasizes agency, control, and abstract systems \cite{lakoff1973language,tannen1990you,giles1991accommodation}.

In the context of causal framing, our results support the notion that gender stereotypes are embedded not just in descriptive language, but also in how cause-effect relationships are constructed. Prior studies in framing theory have shown that the way events and agents are causally linked in discourse can shape public perception and moral evaluation \cite{entman1993framing}. Our findings extend this by showing that such framing is gender-sensitive: Twitter users systematically associate different kinds of consequences with male versus female agents depending on sentiment polarity.

Moreover, the contrast between internal versus external consequences aligns with theories of gendered emotional attribution. Research in psychology has found that women are often perceived as the emotional caretakers or victims in social narratives, whereas men are framed as powerful actors or sources of disruption \cite{eagly1984gender,fiske2018model,rudman2001implicit}. In online settings, these biases can become encoded and amplified at scale. Recent computational studies have shown similar gendered asymmetries in word embeddings, co-occurrence statistics, and machine learning models trained on large corpora \cite{bolukbasi2016man,mehrabi2021survey,stanczak2021survey}.

Thus, the gendered attribution of effects we observe likely reflects a mixture of cultural stereotypes, affective expectations, and narrative roles reinforced through social media discourse. Importantly, these patterns persist even in algorithmically mediated environments, suggesting that social media does not neutralize bias, but can reproduce or even exacerbate longstanding gender tropes.

\section{RQ3 - Gender Attribution and Diffusion Dynamics}

To investigate how gender attribution affects the diffusion of tweets, we analyze the retweet propagation patterns of posts in which the causal subject is identified as either male or female. We focus on two key dimensions of information spread: in-group diffusion, in which retweets go to users belonging to the same social group (community) as the original author; and out-group diffusion, in which  content spreads beyond the original poster’s community into new social communities. This distinction enables us to disentangle the dynamics of localized reinforcement versus network-wide amplification, both of which are critical for understanding how gendered causal language travels within social platforms.

To identify social groups, we construct a retweet network using the full 24-hour Twitter dataset introduced by Pfeffer et al. ~\cite{pfeffer2023just}. This network includes every author–reposter pair during the time window, resulting in a graph with 11,729,159 nodes (users) and 48,648,092 edges (retweet links). The network includes all retweets, regardless of whether the original tweet is part of our narrowed 23-hour causal analysis window. This ensures that we capture full propagation chains and not truncated repost paths.

We apply the Louvain algorithm~\cite{blondel2008fast} for community detection. This modularity-based method partitions the network into clusters by optimizing intra-group density. To account for the stochastic nature of the algorithm, we performed 100 independent runs and retained the dominant clustering pattern. Each detected community is treated as a social group for the purposes of in-group/out-group classification. Prior research shows that community structure in retweet networks often corresponds to ideological alignment, shared identity, or topical affinity~\cite{stieglitz2012political}, making it an appropriate proxy for social grouping in this context.

For each tweet in our dataset that includes a gendered cause (i.e., labeled as Male or Female), we track the subsequent retweets and determine whether each original tweet goes to the same community as the original author (in-group) or to a different one (out-group). This enables us to quantify the extent and breadth of diffusion as a function of gender attribution.

\begin{figure}[tbhp]
\centering
\includegraphics[width=.95\linewidth]{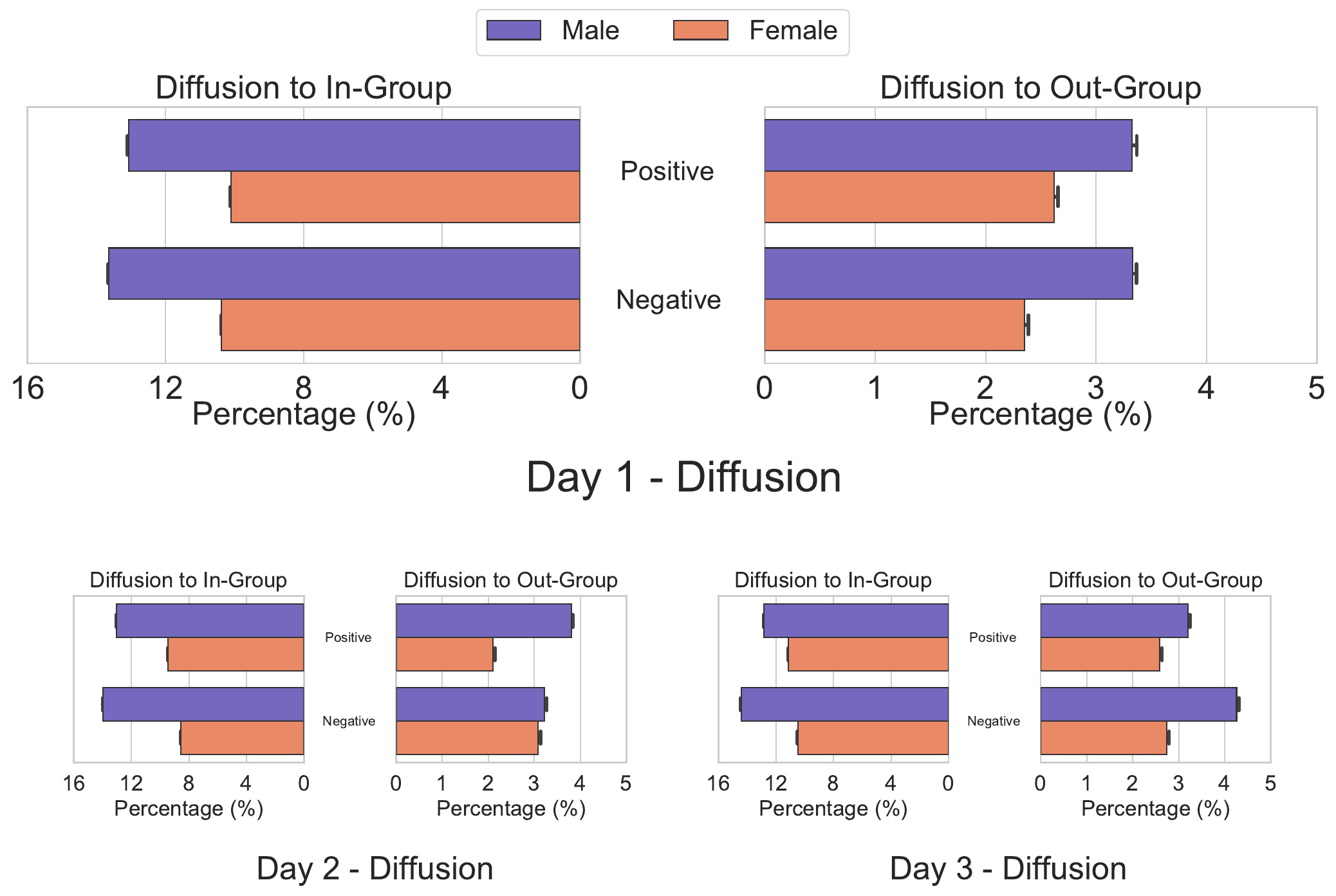}
\caption{Retweet diffusion of gender-attributed causes to in-group and out-group audiences under positive and negative sentiment.}
\label{fig:RQ3}
\end{figure}

Figure~\ref{fig:RQ3} displays the diffusion patterns over the three days in our sample. Each panel separates in-group (left) and out-group (right) diffusion, and compares male- and female-attributed tweets under both positive and negative sentiment conditions.

Across all three days, tweets attributing causes to male subjects exhibit greater diffusion, both within communities (in-group) and across communities (out-group). The disparity is particularly pronounced in in-group diffusion, where male-attributed tweets spread more widely than their female-attributed counterparts, regardless of sentiment. Out-group diffusion, though smaller in magnitude, also favors male-attributed content, particularly under negative sentiment.

These findings contribute to a growing body of research on gendered dynamics in public discourse, particularly how human perception and behavior shape the visibility and credibility of different voices online. Social psychology has long documented that men and women are evaluated differently in terms of authority, trustworthiness, and social influence, especially in public or agentic roles~\cite{eagly1984gender,rudman2001implicit,fiske2018model}. These biases persist in digital settings, where male-authored content is more often granted legitimacy, reposted, or seen as relevant to collective interests~\cite{sotudeh2018gender}.

In our results, male-attributed causal tweets diffuse more widely both within and across communities. This suggests that narratives involving male agents are seen as more worthy of attention or more relevant for sharing—even when the underlying sentiment is negative. These patterns likely stem from cultural expectations that men act as central figures in public, political, or structural events, whereas women are more often associated with personal or interpersonal domains~\cite{eagly1984gender,tannen1990you}.

The disparity is pronounced in in-group diffusion, where users share content from others within their own community. Prior research suggests that social identity and perceived similarity influence what users choose to endorse or propagate~\cite{tajfel1979integrative}. If users implicitly associate male agents with high-status or credible sources—especially in topics linked to power, action, or disruption—this could explain why male-attributed causes gain broader traction.

By contrast, tweets attributing causality to women may be less readily taken up, particularly across community lines. Several studies have found that women’s speech in public settings is more likely to be ignored, interrupted, or devalued~\cite{brescoll2011takes,kricheli2016many}. Even when the content is emotionally resonant, female-attributed narratives may be dismissed as overly personal or lacking broader significance.

Importantly, these differences are not necessarily driven by overt prejudice. Rather, they reflect widespread, often unconscious gender schemas that shape how individuals perceive agency, relevance, and credibility~\cite{heilman2001description,ridgeway2011framed}. In the context of social media, where engagement is rapid and visibility depends heavily on peer endorsement, even subtle perceptual biases can lead to large-scale disparities in diffusion and reach.

Taken together, our results show that gender bias in attribution does not only affect how people write about male and female subjects—it also shapes how these narratives travel. Human tendencies to perceive men as more legitimate causal actors and women as less central or influential appear to extend into the domain of social media sharing. These biases reinforce existing inequalities in who is seen, heard, and believed in public discourse.

\section{Limitations and Ethical Considerations}

The dataset~\cite{pfeffer2023just} used in this study consists of publicly available posts and reposts collected during three complete 24-hour periods. While this dataset provides valuable insight into real-time discourse at scale, it also raises important ethical and methodological considerations. Although the data is public, it may contain personally identifiable information (PII), and some posts may include offensive, harmful, or emotionally charged content. During the data collection period, the platform was known as Twitter, and individual messages were referred to as ``tweets.''  

The dataset~\cite{pfeffer2023just} was released under the Creative Commons Attribution 4.0 International (CC BY 4.0) license and is distributed in accordance with Twitter’s (now X’s) terms of service. Our use complies with these terms and is solely for non-commercial, academic purposes. For the human annotation study used to validate gender attribution, we obtained Institutional Review Board (IRB) approval from the University. Annotators were recruited and compensated through Prolific, and were given clear instructions and screened for quality prior to participation.

We acknowledge that our analysis is constrained to English-language posts and may not capture nuances present in other linguistic or cultural contexts. Even among English speakers, dialect variation, cultural reference, and vernacular expression may influence the effectiveness of our causal span detection pipeline. Certain subgroups or linguistic communities may thus be underrepresented or mischaracterized by our modeling approach.

There is also the potential for misuse of the findings reported in this paper. Quantifying gendered patterns in causal narratives and diffusion could be weaponized to reinforce stereotypes, justify exclusionary design choices, or support discriminatory targeting. For example, one could misinterpret our findings to argue that women are less competent agents in online discourse, or that their narratives are inherently less salient. We strongly caution against such interpretations. The observed disparities are not intrinsic to individuals but are the result of historically rooted social structures and perceptual biases. Our intention is to illuminate these patterns, not to endorse them, and to encourage more equitable attention to how gender shapes narrative authority and visibility in public discourse.

Finally, our study focuses on binary gender attribution (male/female) due to data limitations. Although this decision is methodologically driven, we recognize that it excludes non-binary, transgender, and other gender-diverse voices, whose experiences and linguistic representations are often marginalized. Future work should aim to extend causal analysis frameworks to more inclusive gender representations, with appropriate care and nuance.

\section{Discussion and Conclusion}

Our work presents a comprehensive, large-scale analysis of how gender shapes causal narratives on Twitter. Drawing from millions of posts collected during three full days of platform activity, we examined the affective tone, semantic structure, and diffusion dynamics of causal statements in which the subject was identified as either male or female.

In this work, we find consistent and meaningful asymmetries. Posts attributing causality to female subjects are more likely to be framed with negative sentiment and associated with emotional or relational consequences. In contrast, male-attributed posts are more often linked to positive sentiment and to abstract or systemic outcomes. These differences are not merely rhetorical, they have measurable effects on how content spreads through the social network. Male-attributed causal narratives receive greater in-group and out-group amplification, suggesting higher social salience and legitimacy in the eyes of other users.

Together, these findings reveal how gender biases manifest not just in who is represented or heard online, but in the very structure of explanation and attribution. Who is seen as a cause, what kinds of outcomes they are linked to, and whether those narratives travel, all are shaped by implicit cultural assumptions about gender and agency. Social media, here serves as a mirror for broader social norms that continue to frame male voices as more authoritative and female voices as more affective, private, or marginal.

This study opens several important avenues for future research into gendered discourse, causal framing, and information diffusion on social platforms. Future research should examine platform-level interventions and their effects on the propagation of gendered narratives. For instance, how might changes in visibility algorithms or engagement prompts influence what kinds of gendered causal statements are shared and amplified. Understanding these dynamics would be crucial for designing systems that do not amplify human bias.

%
%
%

\bibliographystyle{splncs04}
\bibliography{mybibliography}







\end{document}